\documentclass[12pt,dvips]{article}

\usepackage{amsmath,amssymb,exscale}
\usepackage{array,multicol}
\usepackage{afterpage,float,flafter}
\usepackage{epsfig,rotating,pifont}
\usepackage{cite}
\@addtoreset{equation}{section} \makeatother

\title{
\vspace{1cm}
\Large\textbf{Brane running and AdS/CFT}
\vspace*{.5cm}
\author{\large \textbf{Michele Redi\footnote{email: redi@pha.jhu.edu}}\\
\emph{
Department of Physics and Astronomy} \\
\emph{Johns Hopkins University} \\
\emph{3400 North Charles St}. \\
\emph{Baltimore, MD 21218-2686}}}

\date{}
\begin{document}
\maketitle \thispagestyle{empty} \vspace*{.5cm}

\begin{abstract}
We extend the formalism introduced in the paper hep-th/0209050 to
compute correlation functions in the AdS/CFT correspondence. We show how the
on-shell action of a scalar field in a compactification of AdS space can
be obtained by flowing in the space of (non-local) theories with
different sizes of the extra-dimension. Our method is relevant in
particular for holographic computations in the Randall-Sundrum
scenario with one or two branes and it allows the inclusion of brane-localized
actions in a systematic way. The method can also be
generalized to other backgrounds and does not rely on explicit
knowledge of the solutions of the wave-equations.
\end{abstract}

\newpage
\renewcommand{\thepage}{\arabic{page}}
\setcounter{page}{1}

\section{Introduction}
The full dynamical content of the AdS/CFT correspondence \cite{maldacena} (for a review \cite{review})
can be summarized by an
equivalence between two path integrals \cite{polyakov}, \cite{witten}:
\begin{equation}
Z[\phi_0]=\displaystyle{\int_{\phi\sim\phi_0}} {\cal D} \phi~\exp(-S[\phi])=\Big<\exp\int_{M_d} d^dx \phi_0 {\cal O}\Big>_{CFT}
\label{adscft}
\end{equation}
Here $\phi$ represents the fields propagating on Anti-de Sitter space (AdS) and ${\cal O}$ are the
dual gauge invariant operators in the Conformal Field Theory (CFT). $Z[\phi_0]$ is the
partition function on (Euclidian) AdS$_{d+1}$, computed with the boundary condition that
the field $\phi$ approaches $\phi_0$ at the boundary of AdS.
The right-hand side of (\ref{adscft}) is the generating functional of the correlation functions
of the dual CFT living on the boundary $M_d$ of AdS$_{d+1}$.
In the limit where classical field theory is a good description of the
physics on AdS (corresponding to the large $N$ limit and large 't Hooft parameter on the CFT side),
\begin{equation}
Z[\phi_{0}]=\exp(-S_{cl}[\phi_0])
\end{equation}
where $S_{cl}[\phi_0]$ is the on-shell value of the classical action for fields that
satisfy the condition $\phi \sim \phi_0$ at the boundary of AdS.
In order to compute $S_{cl}[\phi_0]$ typically one first reconstructs, using the classical
equations of motion, the on-shell value of $\phi$ with the required boundary behavior
and then substitute in the action. This can be done iteratively using
bulk-to-boundary and bulk-to-bulk propagators.

The goal of this paper will be to show how to compute $S_{cl}[\phi_0]$ using
the method of brane running. In principle, no explicit knowledge of bulk-to-boundary
an bulk-to-bulk propagators is required.
This approach was originally proposed in \cite{runrs} and extended in \cite{runrs2} to gauge fields and gravity,
to compute effective field theories of a brane-localized observer in the Randall-Sundrum
scenario with two branes (RS1) \cite{rs}.
As in that paper, we will consider the compactification of AdS on an orbifold $S_1/\mathbb{Z}_2$ with fixed
points at $z_{UV}$ (the ultraviolet brane) and at $z_{IR}$ (the infrared brane), where $z$ is the radial coordinate of AdS.
In the context of the full AdS/CFT correspondence, this geometry should be simply regarded as a regularization:
the $UV$ brane modifies AdS near the boundary and the $IR$ brane cuts the space in the direction of the horizon.
In the RS1 scenario instead, two physical branes placed at the fixed points end the space.
In \cite{runrs} the authors were interested in computing effective theories of an observer localized
on the $IR$ brane, while here we will use the brane running method to compute the on-shell action
as a function of the fields at $z_{UV}$.

The basic idea is that the dependence of the on-shell value of the action on the fields
at $z_{UV}$ can be obtained integrating out bulk degrees of freedom.
Specifically, we consider theories with
different locations of the $IR$ brane. Effective brane actions are generated which guarantee that
$S_{cl}[\phi_0]$ does not change. By moving the $IR$ brane up to $z=z_{UV}$,
by construction $S_{cl}[\phi_0]$ equals the effective $IR$ brane action. Using (\ref{adscft}),
the couplings of the effective $IR$ brane action can be identified with the correlation functions of the CFT.
The couplings on the effective brane "run" in a way
that resembles an ordinary renormalization group (RG) flow. In particular, in the limit where the physical $IR$ brane
is at infinity, corresponding to the RS2 scenario \cite{rs2}, all the brane couplings flow to a non-local fixed point.
Notice that the running of the brane couplings should not be
confused with the RG flow of the observables in
the dual field theory (see for example \cite{kraus1}), since we are keeping the cutoff, represented by the
location of $UV$ brane, fixed.

This formalism appears particularly appropriate for computations in the RS1
scenario. In the holographic interpretation, this model is dual to
a CFT with a cutoff $\Lambda=1/z_{UV}$, where conformal invariance is also broken
in the infrared by the $IR$ brane at the energy scale $1/z_{IR}$.
Since conformal invariance is broken, the couplings on the effective brane
are now not at their fixed points but,
at energies $E>1/z_{IR}$, we can consider this as a small deviation from
conformality. Brane actions can be included with ease, they simply modify the initial
condition for the flow at $z_{IR}$.

The paper is organized as follows. In Section \ref{2point} we present the
general techniques in the case of the $2-$point function of a scalar field and
reproduce well known results.
In Section \ref{interactions} we extend the procedure to $3-$ and
$4-$point functions and consider the general structure of higher-point functions.
In Section \ref{rs} we apply the brane running method to holographic computations
in the RS1 model. Section \ref{conclusions} contains the conclusions.
A more detailed analysis of the $2-$point function is left to
the Appendix.

\section{$2-$point functions}
\label{2point}
We start considering a bulk scalar field propagating in an RS1-type geometry.
We will work in Euclidean AdS$_{d+1}$ with the metric in
Poincar\'e coordinates:
\begin{equation}
ds^2=\frac {L^2} {z^2}(d^dx^2+dz^2)
\end{equation}
where $L$ is the radius of AdS$_{d+1}$.
The radial dimension $z$ is compactified on an orbifold
$S_1/\mathbb{Z}_2$ with fixed points located at $z_{UV}$ (the ultraviolet
brane) and at $z_{IR}$ (the infrared brane). One can think of this
configuration as a regularization of the full AdS, which can
be recovered in the limit $z_{UV} \to 0$ and $z_{IR} \to \infty$.

The Euclidean action of a free scalar of mass $m$ is:
\begin{equation}
S=S_{bulk}+S_{UV}+S_{IR}
\end{equation}
where:
\begin{eqnarray}
S_{bulk}&=&\frac 1 2 \int d^d x \int dz \sqrt{G}(  G^{M N}
\partial_M \phi
\partial_N \phi+ m^2 \phi^2)\nonumber\\
S_{UV}&=&\frac 1 2 \int d^dx \sqrt{g_{UV}}  \phi  \frac {\lambda_{UV}} L
 \phi \nonumber
\\
S_{IR}&=&\frac 1 2\int d^dx \sqrt{g_{IR}}\phi \frac {\lambda_{IR}} L
\phi
\label{quadraticaction}
\end{eqnarray}
$g_{UV}$ and $g_{IR}$ are the induced metrics on the branes.
We will consider $\lambda_{UV}$ and $\lambda_{IR}$ to be derivative couplings. These are
the most general brane actions quadratic in the fields and compatible
with $d-$dimensional Poincar\'e invariance.

Henceforth we take $L=1$. We systematically use the Fourier transform of the field
$\phi$ in the directions parallel to the brane:
\begin{equation}
\phi(x,z)=\int d^dp~ e^{i p x}\phi(p,z)
\end{equation}
In momentum space, the equation of motion is given by:
\begin{equation}
[z^2 \partial_z^2+(1-d)z \partial_z-p^2 z^2 -
m^2-z\delta(z-z_{UV}) \lambda_{UV}(p)-z\delta(z-z_{IR})
\lambda_{IR}(p)]\phi(p,z)=0
\label{massivescalar}
\end{equation}
The brane actions imply boundary conditions on the fields at
$z_{UV}$ and at $z_{IR}$. Assuming that field $\phi$ is even on
the orbifold we have\footnote{Alternatively one can always impose Dirichlet boundary condition. This corresponds
to a different definition of the theory.}:
\begin{equation}
z \frac {\partial} {\partial z}\phi(p,z)\Big|_{z_{IR}}=-\frac
{\lambda_{IR}} 2 \phi(p,z_{IR}) \label{bc}
\end{equation}
Since we are interested in computing $S_{cl}[\phi_0]$,
we consistently impose Dirichlet boundary conditions on the $UV$ brane:
\begin{equation}
\phi(p,z_{UV})=\phi_0(p)
\end{equation}

In the AdS/CFT correspondence, the on-shell value of the action on
AdS as a function of the boundary values of the fields is
the generating functional of the correlation
functions of the gauge invariant operators of the CFT. In
particular, a bulk scalar field of mass $m$ is dual to a scalar
operator of the CFT with conformal dimension $\Delta=d/2 +
\sqrt{(d/2)^2+m^2}$.\footnote{In certain situations the dimension
of the operator in the CFT is given by
$\Delta=d/2-\sqrt{(d/2)^2+m^2}$ \cite{klebanov}. This case can be
obtained by adding appropriate brane actions on the $UV$ brane but
we will not pursue this possibility here.} One popular way to
compute the on-shell action is to consider a Dirichlet boundary
problem at $z=z_{UV}$ \cite{freedman} and use solutions of the wave-equation
(for the two $2-$point function, $S_{cl}[\phi_0]$ depends only
on the asymptotic behavior of the solutions near the boundary;
explicit wave-functions are necessary to compute higher-point functions).

We will now see how the dependence of the on-shell action on the boundary values
of the fields can be
obtained by flowing in the space of theories with different sizes
of the extra-dimension. For applications to the AdS/CFT
correspondence, we are only interested in the effective action seen by an
observer living on the $UV$ brane. We consider a new theory
with an effective $IR$ brane located at $a<z_{IR}$. Requiring that
the original theory and the theory with the effective $IR$ brane at $a$ have precisely the
same solutions in the overlapping region leaves any observable on the
$UV$ brane unchanged. Technically this program is
accomplished including brane actions on the effective $IR$ brane which
provide the same boundary conditions at $a$ as the one implied by
the original brane action at $z_{IR}$. This construction is equivalent to
requiring that $S_{cl}[\phi_0]$ is invariant. Since the
whole information about the CFT is contained in $S_{cl}[\phi_0]$, the AdS theory with
the effective $IR$ brane at $a$ describes the same CFT.

At the quadratic level the necessary effective
brane action is:
\begin{equation}
S_b(a)= \frac {\pi} {a^d} \int d^dp d^dp' \delta(p+p') \phi(p',a)
\lambda_2(a) \phi(p,a) \label{braneaction}
\end{equation}
A few remarks are in order. The coupling $\lambda_2(a)$ has an explicit $a$
dependence and an implicit one from
the momentum (this follows from $d-$dimensional Poincar\'e
invariance). If the original brane action at $z_{IR}$ is local,
i.e. it has an expansion polynomial in $p$, the flow respects such
a property. However, when the physical $IR$ brane is removed to infinity, as
in the full AdS/CFT or in the RS2 model, non-local terms will also be generated. This
is easily understood from a field-theoretic point of view. When AdS
extends to infinity we have a continuum of $d-$dimensional
particles which generates branch cuts in the effective action seen by an
observer located on the $UV$ brane.

These non-local terms are
crucial since they reproduce the correlation functions of the
CFT at non-coincident points. Therefore we consider brane actions
which contain local and non-local terms:
\begin{equation}
\lambda_2(a)=\lambda^{loc.}_2(a)+\lambda^{n.loc.}_2(a)
\end{equation}
$\lambda_2^{loc.}$ has a polynomial expansion in momentum:
\begin{equation}
\lambda^{loc.}_2(a)=\sum_j \lambda_2^{(2j)}(a)(g_a^{\mu\nu} p_\mu
p_\nu)^j
\end{equation}
where $g_a^{\mu\nu}$ is the induced metric at $z=a$.
The precise form of $\lambda_2^{n.loc.}$ depends on the value of
the bulk mass.

The flow of the brane couplings, as a function of the
radial coordinate $a$, can be cast in a form that resembles an ordinary RG
flow.
In the theory with the brane action (\ref{braneaction}),
the boundary condition at $a$ is:
\begin{equation}
a \frac {\partial} {\partial a} \phi(p,a)=-\frac {\lambda_2(a)} 2
\phi(p,a) \label{bca2}
\end{equation}
In order to compute the differential flow of $\lambda_2$, we take the
logarithmic derivative with respect to $a$ of (\ref{bca2}). Since by hypothesis the
solutions remain unchanged in the overlapping region, we use the bulk
equation of motion to eliminate the second derivatives of the
fields. Requiring that the resulting equation is satisfied for any
value of $\phi$, we obtain the flow equation of $\lambda_2$:
\begin{equation}
a \frac d {da} \lambda_2= \frac {\lambda_2^2} 2+d \lambda_2 -2 m^2
-2 a^2 p^2 \label{flowequation}
\end{equation}
This is the main equation of this section. It contains the same information
about the CFT as does the original action.
One can integrate this equation with the boundary condition:
\begin{equation}
\lambda_2(z_{IR})=\lambda_{IR}
\end{equation}
to obtain the couplings of the effective brane at $a$.

Solving (\ref{flowequation}) allows one to compute the on-shell value
of the action in terms of the values of the fields at $z_{UV}$.
Explicitly, "running" the effective brane all the way down to the
$UV$ brane, $S_{cl}[\phi_0]$ is simply given by the sum of the $UV$ and the effective
brane actions, because the bulk contribution goes to zero.
Assuming for simplicity that $\lambda_{UV}=0$, we have from
(\ref{braneaction}):
\begin{equation}
S_{cl}[\phi_0]=S_b(z_{UV})= \frac {\pi} {z_{UV}^{d}} \int d^dp
\delta(p+p') \phi_0(p') \lambda_2(z_{UV}) \phi_0(p)
\label{scl2}
\end{equation}

The flow equation (\ref{flowequation}) has a fixed point in the sense $a\partial_a
\lambda_2(a)=0$ (that is, the coupling depends on $a$ only through
the dependence on momentum):
\begin{equation}
\lambda^{f.p.}_2(pa)=(2\nu-d)+2 p a \frac {K_{\nu-1}(pa)}
{K_\nu(pa)}
\label{fixedpoint}
\end{equation}
where $K_\nu$ is the modified Bessel function of order
$\nu=\sqrt{(d/2)^2+m^2}$. In the limit where the $IR$ brane is
at infinity, the physics on the $UV$ brane does not depend
on the $IR$ brane. In this case (see Appendix) the coupling $\lambda_2(a)$ is a
function of $pa$ only: it is at its fixed point. In other
words, choosing the brane coupling at the fixed point (\ref{fixedpoint})
is equivalent,
from the point of view of an observer living on the $UV$ brane, to
removing the $IR$ brane to infinity. In the Appendix we show that
the flow equation (\ref{flowequation}) has in general a line of fixed points, but these
correspond to solutions of the wave-equation which are not
regular at infinity, so they are rejected. One can check that, in the limit
$z_{IR}\to \infty$ and zero brane actions, $\lambda_2(a)$ approaches
the fixed point (\ref{fixedpoint}).

According to the AdS/CFT prescription (\ref{adscft}), $\lambda_2(z_{UV})$ is, up to a normalization, the
$2-$point function of a scalar operator of the CFT of dimension $\Delta=d/2+\nu$
(in the large $N$ limit and strong 't Hooft parameter).
In fact, with $\lambda_2$ at the fixed point (\ref{fixedpoint}), the brane
action evaluated at $z_{UV}$ coincides up to a factor of two
(coming from working in the orbifold covering space) with the
standard AdS/CFT on-shell action computed, for example, in
\cite{freedman}. The expansion of (\ref{fixedpoint}) for integer $\nu$ is:
\begin{equation}
\lambda_2^{f.p.}(pa)=2\nu-d+\frac {p^2a^2}
{\nu-1}+...+\tilde{\lambda}_2^{(2\nu)}(pa)^{2\nu}\log{pa}+...
\end{equation}
where the coefficient $\tilde{\lambda}_2^{(2\nu)}$ depends on $\nu$ only.
For non-integer $\nu$ one finds a similar expansion with first non-analytic
term $(pa)^{2\nu}$. Notice that, as promised, the brane action
necessary to reproduce the infinite space limit is non-local. In AdS/CFT computations
one then proceeds taking the limit
$z_{UV}\to 0$ in (\ref{scl2}). The divergent local terms in $S_{cl}[\phi_0]$ need to be
renormalized away \cite{skenderis}. The non-local term $p^{2\nu} \log p$ reproduces the
$2-$point function at non-coincident points of
a scalar operator with conformal dimension $\Delta=d/2+\nu$.

\section{Interactions}
\label{interactions}
Higher-point functions can be computed in a very similar way to
the $2-$point functions, by integrating out bulk degrees of
freedom.
\subsection{$3-$point functions}
\label{3point}
As an example we will consider the case of cubic
interactions. The bulk action in momentum space contains:
\begin{equation}
S_{int}= 2\pi \sigma^{ijk} \int \sqrt{G} dz d^dp d^dq d^dk \delta(p+q+k)
\phi_i(p,z) \phi_j(q,z) \phi_k(k,z)
\end{equation}
where $\phi_i(x,z)$ are scalar fields of mass $m_i$ and
$\sigma^{ijk}$ is symmetric in the indexes.
The bulk equation of motion reads:
\begin{equation}
[z^2 \partial_z^2+(1-d)z \partial_z-z^2p^2-m^2)]
\phi_i(p,z)-3 \sigma^{ijk} \int d^dq \phi_j(p-q,z)\phi_k(q,z)=0
\end{equation}
In this case the effective brane action also contains cubic and higher order
terms:
\begin{equation}
S_b(a) \supset \frac {2\pi} {a^d} \int d^dp d^dq d^dk \delta(p+q+k)
\lambda_3^{ijk}(p,q,k) \phi_i(p,a) \phi_j(q,a) \phi_k(k,a)
\end{equation}
The boundary condition at $a$ becomes:
\begin{equation}
a\frac {\partial} {\partial a} \phi_i(p,a)=-\frac {\lambda_2^i(p)}
2 \phi_i(p,a)- \frac 3 2 \int d^dq \lambda_3^{ijk}(-p,p-q,q)
\phi_j(p-q,a) \phi_k(q,a) \label{bca3}
\end{equation}
where $\lambda_2^i$ is the quadratic brane coupling of the field
$\phi_i$.

To compute the flow equation for the coupling $\lambda_3^{ijk}$,
we proceed by taking the
logarithmic derivative of (\ref{bca3}), as in the case of the $2-$point function. We then use the bulk
equation of motion and require that the equation is satisfied
for any value of the fields. With a cubic brane action this can be
done up to order $\phi^2$ in the fields. Of course,
this is because a cubic interaction generates higher-point functions as well. The important point is that, at the
classical level, lower order point functions do not mix with
higher orders ones, so the flow of the coupling $\lambda_i$ depends
only on the couplings $\lambda_j$ with $j\le i$.

The equation for $\lambda_3^{ijk}$ is:
\begin{equation}
a \frac {d} {da} \lambda_3^{ijk}(-p,p-q,q) =-2 \sigma^{ijk}+d
\lambda_3^{ijk}(-p,p-q,q)+\frac 1 2 \lambda_3^{ijk}
(-p,p-q,q)[\lambda_2^i(p)+\lambda_2^j(p-q)+\lambda_2^k(q)]
\label{threepoint}
\end{equation}
While this equation can be difficult to solve in general, it can
be solved perturbatively expanding in powers of momenta, as
explained in the Appendix for the case of the $2-$point function.

In the infinite AdS case, each brane
coupling should be set at the fixed point, so they do not depend explicitly
on $a$. One can check that the fixed point in this case is given by:
\begin{equation}
\lambda_3^{ijk}(-p,p-q,q)=2 \sigma^{ijk} a^d \int_a^\infty dz \frac
{1} {z^{d+1}}\Big(\frac {z} {a} \Big)^{\frac {3d} {2}} \frac
{K_{\nu_i}(-pz)K_{\nu_j}((p-q)z)K_{\nu_k}(q z)}
{K_{\nu_i}(-pa)K_{\nu_j}((p-q)a)K_{\nu_k}(q a)}
\label{fp3}
\end{equation}
Notice that, since the $\beta-$function for $\lambda_3^{ijk}$ is linear in $\lambda_3^{ijk}$,
the fixed point is completely determined by $\lambda_2^{f.p.}$. The fixed point (\ref{fp3})
is just the $3-$point function in the limit $z_{IR} \to \infty$,
computed with bulk-to-boundary propagators \cite{victoria}.

\subsection{Higher-point functions}
\label{4point}
The method of brane running is completely general and can be extended to arbitrary order
point functions. The computation of $4-$ and higher-point functions is complicated in the AdS/CFT correspondence
because it involves bulk-to-bulk AdS propagators. In our case,
the structure of the differential equations for the
higher-point functions is similar to the case of the $3-$point
function and can be readily understood in a diagrammatic way.

In order to compute the  $4-$point function we need to include a
quartic brane action:
\begin{equation}
S_b(a) \supset \frac 1 {a^{d}} \int d^dx
\lambda_4^{ijkl}\phi_i\phi_j\phi_k\phi_l
\end{equation}
Following the same steps as for the $3-$point function, one
obtains:
\begin{eqnarray}
&&a\frac d {da}\lambda_4^{ijkl}=d \lambda_4^{ijkl}+\frac {\lambda_4^{ijkl}}
2[\lambda^i_2(p)+\lambda^j_2(p-q-k)+\lambda^k_2(q)
+\lambda^l_2(k)]+\nonumber \\ &&+\frac
9 8 [\lambda_3^{imk}(-p,p-q,q)\lambda_3^{mjl}(q-p,p-q-k,k)+\lambda_3^{ijm}(-p,p-q-k,q+k)
\lambda_3^{mlk}(-q-k,q,k)]\nonumber \\
\label{4fp}
\end{eqnarray}
In the previous equation, the momentum dependence of the coupling is $\lambda_4(-p,p-q-k,q,k)$. The fixed point
of (\ref{4fp}) is, up to normalization, the amplitude for the $4-$point
function in the AdS/CFT which can be expressed in terms of bulk-to-bulk
and bulk-to-boundary propagators.

We now analyze the general structure of the $\beta-$functions for
the couplings. We consider in particular the $4-$point
function but the same considerations hold for higher-point
functions. Schematically, the diagrams contributing to the $4-$point
function when brane actions are included are represented in
the figure below.
\begin{figure}[ht]
\centerline{\epsfxsize=1.5in \epsfbox{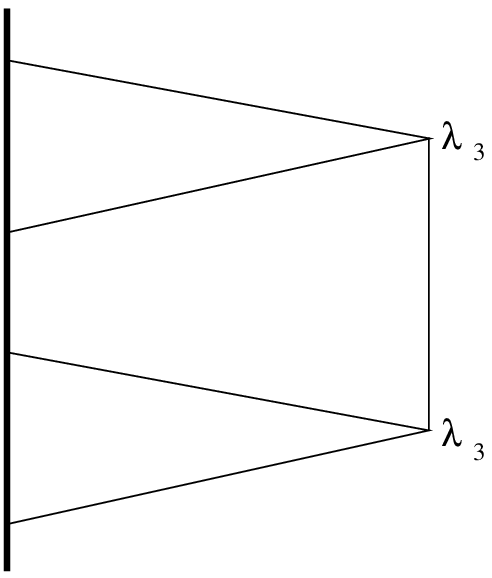} \hspace {1.5cm}
\epsfxsize=1.5in \epsfbox{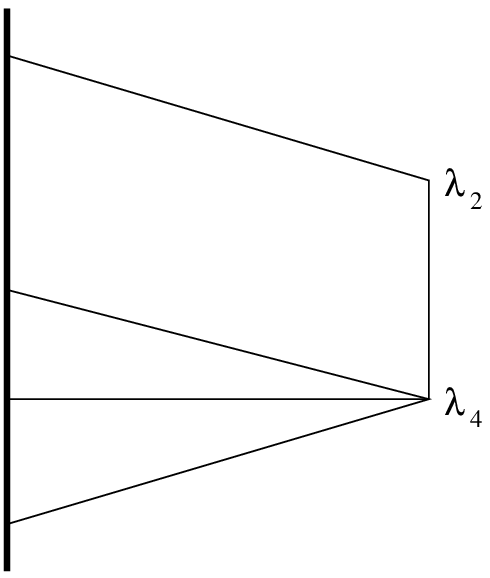}  \hspace{1.5cm}
\epsfxsize=1.35in \epsfbox{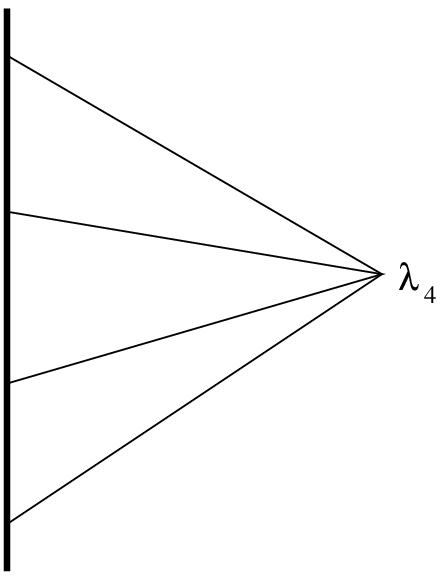}} \caption{{\it Diagrams
contributing to the "running" of $\lambda_4$ at the classical level.
The external legs are located on the $UV$ brane and the interactions
take place on the effective brane located at $a$.}}
\end{figure}

To move the brane from $a$ to $a-da$, we need to provide
counter-terms for the degrees of freedom which have been integrated
out. As a consequence, the infinitesimal variation of $\lambda_4$ contains
a contribution proportional to $\lambda_3^2$, a contribution
proportional to $\lambda_2\lambda_4$ as well as a contribution
linear in $\lambda_4$. A quartic bulk interaction would act as a
source in the $\beta-$function of $\lambda_4$.
The dependence on the momenta can be worked out as in ordinary Feynman diagrams.

\section{Applications}
\label{rs}
In this section we apply the brane running method
to the Randall-Sundrum scenario
\cite{rs}. We take $d=4$.

In the full AdS/CFT correspondence, a theory on AdS is
equivalent to a pure conformal field theory living on the
boundary. In the previous analysis the $UV$ brane and the $IR$ brane
should be understood as regulators of the theory so the physical
quantities are extracted in the limit $z_{UV}\to 0$ and $z_{IR}\to
\infty$. As we have argued, the second limit (corresponding to the geometry
of the RS2 model) implies that all the
couplings on the effective brane are at their fixed points.
In the RS1 scenario, five dimensional AdS
is truncated by two physical branes parallel to the boundary,
placed at the fixed points of an orbifold. Since small values of
the coordinate $z$ correspond to the ultraviolet of the CFT, the presence of the
$UV$ brane represents a cutoff applied to the CFT. On the other
hand, the $IR$ brane modifies the space only at large $z$ so naively it
corresponds, in the holographic picture, to breaking
(spontaneously) conformal invariance in the infrared, at a scale
$1/z_{IR}$. It should be clear from the present paper that this
conclusion may not hold when brane actions are included: theories with
different size of the extra-dimension can have the same CFT dual.
Since for phenomenological applications one takes
$z_{IR}/z_{UV} \sim 10^{15}$, the theory should appear
approximately conformal for a large range of energies, namely
$1/z_{IR}<E<1/z_{UV}$.

To any order in the fields, the dependence of the action on the
boundary values of the fields can be computed integrating the flow
equations presented in the previous sections. This can be done
perturbatively in $p$ as shown in the Appendix for the $2-$point
function. In particular it is straightforward, even for higher-point functions,
to compute the zero momentum dependence of the
coupling $\lambda_n$. The leading dependence on the external
momentum is also not hard to compute. Since the $IR$ brane breaks
conformal invariance in the infrared, the brane couplings are not at
a fixed point, but acquire an explicit dependence on the
scale $a$.

In the case of the $2-$point function, equation
(\ref{flowequation}) can be solved explicitly, the solution being:
\begin{equation}
\lambda_2(a)=2\nu-4 +2 p a\frac {K_{\nu-1}(p a)+C~ I_{\nu-1}(pa)} {K_\nu(p a)- C~ I_\nu(p
a)}
\end{equation}
The integration constant $C$ is determined by the boundary condition
at the $IR$ brane, namely $\lambda_2(z_{IR})=\lambda_{IR}$. This allows
the inclusion in a simple and systematic way of arbitrary brane
actions. For example, for a massless scalar and $\lambda_{IR}=0$, corresponding to
ordinary Neumann boundary condition, the result is:
\begin{equation}
\lambda_2(a)=2 p a~ \frac {I_1(p z_{IR})K_1(p a)-I_1(p a)K_1(p
z_{IR})} {I_2(p a) K_1(p z_{IR})+I_1(p z_{IR})K_2(p a)}
\label{2pirs}
\end{equation}
For $a=z_{UV}$ this is just the analytic continuation to Euclidean AdS of the
inverse of the brane-to-brane propagator of a massless scalar
computed in \cite{nolte}. In general,
$\lambda_2(z_{UV})$ is the inverse of the brane-to-brane propagator, so
the zeros in $p$ of $\lambda_2(z_{UV})$ correspond
to the masses of the Kaluza-Klein (KK) modes.

For $pz_{IR}<1$ one can expand $\lambda_2(a)$ in powers of $p$. The
expansion is now completely polynomial in momentum. Technically
this happens because if the boundary condition is local, this
property will be preserved by the flow. From the CFT point of view,
this fact is simply the statement that below the scale $1/z_{IR}$,
conformal symmetry is completely broken, and in particular, a mass
gap is generated. On the AdS side we have a discrete set of
$4D$ particles whose mass scale is set by $1/z_{IR}$.\footnote{For the case $-d^2/4\le m^2<0$,
i.e. tachyons satisfying the lower bound on the masses of scalars in AdS$_{d+1}$,
the KK spectrum also contains a single tachyon whose mass squared is $O(-1/z_{UV}^2)$ \cite{redi}.}
This discrete set of particles gives rise
to isolated poles in the propagator but no branch cuts. Therefore,
at energies below the masses of the KK modes, this system should be
describable in terms of a local action.

At energies above
$1/z_{IR}$, conformal invariance is approximately restored. For
$pz_{UV}<1$, i.e. energies below the cutoff of the CFT, we can
expand $\lambda_2(z_{UV})$ around $z_{UV}=0$. In the case of the massless scalar (\ref{2pirs}) we have:
\begin{equation}
\lambda_2(z_{UV})=p^2 z_{UV}^2+\frac 1 2 (\gamma+\log\frac
{pz_{UV}} 2) p^4 z_{UV}^4 - \frac {K_1(p z_{IR})} {2 I_1(p
z_{IR})} p^4 z_{UV}^4+ O(z_{UV}^6)
\end{equation}
In the limit $z_{IR} \to \infty$ we recover the expansion of the
fixed point. This should be expected since in this limit all
the couplings flow to their fixed points. Using the asymptotic expansion
of the Bessel functions, it is easy to see that the deviation
from $\lambda_2^{f.p.}$ becomes exponentially small at energies greater that $1/z_{IR}$. Since for
large momenta the presence of the $IR$ brane represents a small
deviation from conformality, all the brane couplings are close to the
fixed point. In this regime it is useful to simplify the flow equation
expanding to linear order around $\lambda_2^{f.p.}$:
\begin{equation}
a\frac d {da} \alpha_2=\alpha_2\lambda_2^{f.p.}+ d
\alpha_2
\end{equation}
where $\alpha_2=\lambda_2-\lambda_2^{f.p.}$. To leading order, for $pa>1$, the
solution of the previous equation is simply $\alpha_2=C~a~e^{2 p a}$, independently on $\nu$.
As we move the effective brane to smaller $a$, $\alpha_2(a)$ becomes exponentially suppressed.
Similar manipulations can be done for higher-point functions.

Localized fields can be naturally included in this
formalism. This case is important for phenomenological
applications. For example, in the original proposal by Randall and
Sundrum, in order to solve the hierarchy problem, the Standard
Model fields were localized on the $IR$ brane. For fields
localized on the $UV$ brane, it is obvious that their action should
be added to the effective action. From the CFT point of
view, fields living on the $UV$ brane are interpreted as fields
external to the CFT and coupled to gauge invariant operators
of the CFT so it is natural to add the action of the CFT and that
of the external fields. In the case of fields localized on the $IR$ brane,
or in the bulk, the analysis is more intricate because
these fields couple only indirectly to the $UV$ brane. As an
example, following \cite{victoria}, we consider a free scalar
field coupled to an external source $j$ localized at $z=z_0$. In $d+1$ dimensions, the
AdS action contains:
\begin{equation}
S\supset \int d^dx dz \sqrt{G} j \phi z \delta(z-z_0)
\end{equation}
Consistency requires the brane action:
\begin{equation}
S_b(a)= \frac 1 {a^{d}}\int d^dx \lambda_1 \phi +\frac 1 {2
a^{d}}\int d^dx \phi \lambda_2 \phi
\end{equation}
The running of $\lambda_1$ is given by:
\begin{equation}
a\frac d {da} \lambda_1=d\lambda_1+\frac {\lambda_1\lambda_2} 2
\label{onepoint}
\end{equation}
while the $\beta-$function of $\lambda_2$ remains unchanged. Notice that only
in this particular case a coupling depends on higher order ones.\footnote{If bulk
interactions are included the $\beta-$function of the coupling
$\lambda_i$ will depend on the couplings up to $\lambda_{i+1}$.}
For a localized source in the case where the $IR$ brane is at infinity,
$\lambda_2$ will still be
at its fixed point (\ref{fixedpoint}). It is not hard to check that the
solution of (\ref{onepoint}) is:
\begin{equation}
\lambda_1=(\frac a {z_0})^{d/2}\frac {K_\nu(pz_0)} {K_\nu(pa)}
j(p)
\end{equation}
A tadpole for the field $\phi$ is generated in the effective field
theory of an observer on the $UV$ brane. On the CFT side, this
corresponds to an expectation value for the dual operator.

\section{Conclusions and Outlook}
\label{conclusions}
In this paper we have shown how to compute the value of the on-shell
AdS action of a scalar field by flowing in the space of
theories with a different size for the extra-dimension. By
integrating out bulk degrees of freedom, we generate effective
brane actions which reproduce exactly the same physics in the
restricted region. To any order in the fields, the dependence of the on-shell action on the
boundary values of the fields at $z_{UV}$ is obtained by solving the flow
equations for the couplings on the effective brane.
In principle, this procedure does not require the
knowledge of the bulk-to-boundary and bulk-to-bulk propagators.
The values of the couplings are related in a simple way to the
correlation functions of the CFT. In the case of infinite AdS or RS2, the
brane couplings are at their fixed points, corresponding to an unbroken
CFT.

This formalism is particularly appropriate for holographic
computations in the RS1 model. Here the
couplings on the effective brane are not at the fixed points but
acquire an explicit dependence on the radial coordinate. At energies much
above the scale set by the $IR$ brane, corresponding to the scale
where conformal invariance is broken, the couplings are very close
to the fixed points so we can simplify the flow equations.

Gauge fields and fluctuations of the metric can be included
in a very similar way. The relevant steps are presented in
\cite{runrs2}, where the brane running is used to compute
the effective field theory of an observer living on the $IR$
brane. The extension to fermionic fields is non-trivial because the
wave-equation is a first order differential equation in this case.

More in general, the method of brane running can be used in geometries
which contain a Poincar\'e invariant subspace.
The structure of the $\beta-$functions for the
couplings would be similar to the one in the present paper. In
this case though, the brane couplings, even in the infinite space
limit, will not be at a fixed point.

\section*{Acknowledgments}
I would like to thank Kaustubh Agashe and Jon Bagger for reading the manuscript.
This work was supported in part by the U.S. National Science Foundation,
grant NSF-PHY-9970781 and NSF-PHY-0099468.

\appendix
\section*{Computation of the $2-$point function}
\label{appendix}
In this Appendix we give a self-contained derivation of the flow
equation for the coupling $\lambda_2$ and carefully study the
solutions.

The boundary condition at $a$ determined by the brane
action is:
\begin{equation}
z \partial_z\phi(p,z)\Big|_a=-\frac {\lambda_2(a)} 2 \phi(p,a)
\label{bcapp}
\end{equation}
The bulk equation of motion is:
\begin{equation}
(z^2 \partial_z^2+(1-d) z\partial_z-p^2 z^2-m^2)\phi(p,z)=0
\label{eqmapp}
\end{equation}
We are interested in the differential flow of the coupling
$\lambda_2$. By assumption the solution of the equations of
motion do not depend on the location of the brane. This physical
condition determines a (classical) flow for $\lambda_2$. Taking
the logarithmic derivative of (\ref{bcapp}) with respect to $a$ we
obtain:
\begin{equation}
(a \frac {d} {da} \lambda_2+ \lambda_2 a\partial_a+ 2
a\partial_a+2 a^2\partial_a^2)\phi(p,a)=0
\end{equation}
Using the bulk equation of motion to eliminate the second
derivative and requiring that this new equation is satisfied for
any value of the field $\phi(p,a)$, we find the flow equation:
\begin{equation}
a \frac {d} {da} \lambda_2=\frac {\lambda_2^2} 2 + d \lambda_2-2
m^2- 2 p^2 a^2 \label{flowapp}
\end{equation}
The solutions of this equation are functions of $pa$. The boundary
condition at $z_{IR}$ introduces a dependence on $p z_{IR}$. A quick way
to find the solutions
of (\ref{flowapp}) is to use explicit wave-functions and substitute in (\ref{bcapp}).
The solution of the bulk equation of motion is:
\begin{equation}
\phi(p,z)=z^{d/2}(A K_\nu(p z)+B I_\nu(pz))
\end{equation}
where $I_\nu$ and $K_\nu$ are Bessel's functions of order $\nu=\sqrt{(d/2)^2+m^2}$.
Plugging in (\ref{bcapp}) and using properties of the Bessel's functions we find:
\begin{equation}
\lambda_2(a)=2 \nu-d+ 2pa \frac {A K_{\nu-1}(p a)-B I_{\nu-1}(p
a)} {A K_{\nu}(p a)+B I_{\nu}(p a)} \label{fpapp}
\end{equation}
The coefficients $A$ and $B$ have a dependence on $pz_{IR}$ determined by the $IR$ brane.
Particularly important for the
applications to the AdS/CFT are the solutions of this equation
which depend on $p a$ only. We call these solutions fixed points
because the couplings in the expansion of $\lambda_2$ are
independent on $a$ in this case.
From (\ref{fpapp}) we see that the flow equation has a line of fixed points depending on the ratio
$A/B$ (with $A/B$ independent of $p$). Setting $\lambda_2$ at
a fixed point is physically
equivalent, for an observer living on the $UV$ brane, to removing the $IR$ brane to infinity.
Regularity at infinity selects the wave-function $K_\nu$ so the relevant
fixed point corresponds to $B=0$.

For illustrative purposes we now show an alternative method to
compute the fixed points of (\ref{flowapp}). The same technique could be used to solve the
equations for the higher order couplings or to study the $2-$point
functions in backgrounds which are only asymptotically AdS.

Equation (\ref{flowapp}) can be solved iteratively expanding in
powers of $p$. Notice that for the fixed points the momentum
expansion is the same as the expansion around $a=0$, that is a near boundary
analysis. For simplicity, we will consider the case of $\nu$
integer which is the most common in the string theory literature. We
start with the following ansatz for the solution:
\begin{equation}
\lambda_2(a)=\lambda_2^{(0)}(a)+ \lambda_2^{(2)}(a)
(pa)^2+.....+\lambda_2^{(2\nu)}(a)(pa)^{2\nu}+ ....
\end{equation}
Substituting in the flow equation we obtain a set of coupled
differential equations for the coefficients $\lambda_2^{(2i)}$:
\begin{eqnarray}
a\partial_a \lambda_2^{(0)}&=&\frac {(\lambda_2^{(0)})^2} 2 + d
\lambda_2^{(0)}-2 m^2 \nonumber \\ a\partial_a
\lambda_2^{(2)}&=&\lambda_2^{(2)}\lambda_2^{(0)}+(d-2)\lambda_2^{(2)}-2
\nonumber \\a\partial_a
\lambda_2^{(2n)}&=&\lambda_2^{(2n)}(\lambda_2^{(0)}+d-2n)+\frac 1
2 \sum_{i+j=n} \lambda^{(2i)} \lambda^{(2j)} \label{flowexpapp}
\end{eqnarray}
For the fixed points all the derivatives are equal to zero so we
have a set of algebraic equations. The solution of the first
equation in (\ref{flowexpapp}) is:
\begin{equation}
\lambda_2^{(0)}=-d \pm 2 \nu
\end{equation}
The solution with the $-$ sign corresponds to the fixed point with
$I_\nu$ only. This solution has an expansion purely polynomial in
momentum. In position space this generates only contact
interactions. With the choice of the $+$ sign we can solve
(\ref{flowexpapp}) iteratively only up to the order $2\nu$. This
equation is inconsistent. In order to find a solution we need to
include a term $\tilde{\lambda}_2^{(2\nu)} (qa)^{2\nu}\log qa$ in
the expansion of $\lambda_2$. Now the equation $(2\nu)$ determines
the coefficient $\tilde{\lambda}_2^{(2\nu)}$ while
$\lambda_2^{(2\nu)}$ remains unconstrained. The fact that
$\lambda_2^{(2\nu)}$ is undetermined is of course related to possibility of
choosing $A/B$ in (\ref{fpapp}).

\end{document}